# Methods to induce perpendicular magnetic anisotropy in full-Heusler $Co_2FeSi$ thin layers in a magnetic tunnel junction structure

Koki Shinohara, Takahiro Suzuki, Yota Takamura, Shigeki Nakagawa[*]

Electrical and Electronics Engineering, School of Engineering,
Tokyo Institute of Technology,
2-12-1 Ookayama, Meguro-ku, Tokyo 152–8552, Japan

[*]The corresponding author
Shigeki Nakagawa
nakagawa@ee.e.titech.ac.jp*

**ABSTRACT**

In this study, to obtain perpendicular magnetic tunnel junctions (p-MTJs) using half-metallic ferromagnets (HMFs), several methods were developed to induce perpendicular magnetic anisotropy (PMA) in full-Heusler $Co_2FeSi$ (CFS) alloy thin layers in an MTJ multilayer composed of a layered CFS/MgO/CFS structure. Oxygen exposure at 2.0 Pa for 10 min after deposition of the bottom CFS layer was effective for obtaining PMA in the CFS layer. One of the reasons for the PMA is the formation of nearly ideal CFS/MgO interfaces due to oxygen exposure before the deposition of the MgO layer. The annealing process was effective for obtaining PMA in the top CFS layer capped with a Pd layer. PMA was clearly observed in the top CFS layer of a Cr(40 nm)/Pd(50 nm)/bottom CFS(0.6 nm)/MgO(2.0 nm)/top CFS(0.6 nm)/Pd(10 nm) multilayer, where the top CFS and Pd thin films were deposited at RT and subsequently annealed at 300°C. In addition to the continuous layer growth of the films, the crystalline orientation alignment at the top CFS/Pd interface probably attributes to the origin of PMA at the top CFS layer.

**INTRODUCTION**

Perpendicular-anisotropy magnetic tunnel junctions (p-MTJs)[1–3] have attracted considerable attention for their potential application in spin-transfer torque magnetoresistive random access memories (STT-MRAMs). Various ferromagnetic thin films with perpendicular magnetic anisotropy (PMA), such as (Co, Fe)-Pt alloys[4,5] and Co-Pt(Pd) superlattices[6,7], have been studied to be used in p-MTJs. Recently, p-MTJs comprising CoFeB/MgO/CoFeB tri-layered structures[8] were developed, exhibiting a high tunnel magnetoresistance (TMR) ratio above 120%. This has accelerated the development of STT-MRAMs. However, CoFeB does not have a very high spin polarization ratio, $P$, which is approximately 65%[9,10]. Therefore, investigation of PMA materials with higher $P$ values is required for next generation STT-MRAMs with PMA.

Half-metallic ferromagnets (HMFs)[11–13] are the highest spin polarized materials with a perfect $P$ of 100%. Many Co-based full-Heusler alloys, such as $Co_2MnSi$ (CMS), $Co_2FeSi$ (CFS), and $Co_2FeSi_xAl_{1-x}$, have been theoretically predicted[12,13] to be HMFs, and some have experimentally exhibited half-metallicity[14,15]. Despite the excellent half-metallic behavior of CMS at low temperature, the TMR of CMS-MTJs considerably decreases as increasing the temperature. However, CFS is expected to show less dependence on TMR signals[16–18]. Similar to other full-Heusler alloys, CFS has a low damping constant $\alpha$ of 0.008[19], which can reduce the critical current density for current-induced magnetization switching. However, because of their highly symmetric cubic crystal structure, full-Heusler CFS alloy thin films[20] do not exhibit PMA.

Recently, interfacial PMA induced by the Fe–O hybrid orbital in Fe alloy/MgO interfaces has been extensively studied[21]. This interfacial PMA has also been reported in full-Heusler $Co_2FeAl$ alloy thin films[22], although it is intrinsically non-half-metallic. In our previous report[16], we successfully observed PMA in CFS/MgO bilayered structures, which potentially have the half-metallic feature. Since the MgO capping layer can be used as a tunnel barrier for p-MTJs, this CFS layer with PMA is applicable to the bottom ferromagnetic electrode. Nevertheless, a CFS layer formed on an MgO layer, which can be applied to the top electrode, did not exhibit PMA.

In this report, we developed perpendicularly magnetized CFS layers formed on MgO layers using several techniques, and we successfully fabricated p-MTJ structures with two perpendicularly magnetized CFS layers.

**EXPERIMETNAL METHODS**

All the samples were formed on Cr(50 nm)/Pd(40 nm)/MgO (001) single crystalline substrates using a facing targets sputtering (FTS) system with a base pressure of 3.0 × 10$^{-5}$ Pa. Cr and Pd buffer layers were deposited in 40-nm-thick and 50-nm-thick layers, respectively, in this order using a DC FTS at a substrate temperature $T_S$ of RT. Further, a 0.6-nm-thick CFS layer was deposited at $T_S$ = 300°C with the DC FTS technique using alloy targets. After cooling to RT, the CFS layer was exposed to an oxygen atmosphere at a pressure of 2.0 Pa for 10 min. Subsequently, a 2-nm-thick MgO layer was deposited via RF FTS and then the stack structure was capped with Ta for samples that only had a bottom magnetic layer. The stack structure is shown in Fig. 1(a). For samples with both magnetic electrodes, a 0.6-nm-thick CFS layer was deposited on the MgO layer and then capped with a 10-nm-thick Pd layer.

The magnetic properties of the samples were measured using a vibrating sample magnetometer, and the magnetic anisotropy energies were estimated from the gap between the magnetization curves for the perpendicular and in-plane magnetic fields. The surface morphology was characterized using an atomic force microscope (AFM) and the crystallographic structure of the stacks was analyzed by using the x-ray diffraction (XRD) method. The CFS films deposited on the Cr/Pd buffer layers at $T_s \geq$ 300 °C showed $L2_1$-ordered structure with (001) orientation. In addition, the chemical composition of the sputtered CFS films was evaluated to be $Co_{51.0}Fe_{24.9}Si_{24.1}$ by an electron probe micro analyzer.

**RESULTS AND DISCUSSION**

Figures 1(b) and 1(c) compare the *M–H* characteristics of the Cr/Pd/CFS/MgO/Ta samples with and without oxygen exposure, respectively. As shown in Fig. 1(b), the easy magnetization direction for the sample without oxygen exposure was parallel to the film plane, indicating that in-plane magnetic anisotropy of the film is preferable even though a CFS/MgO interface was formed. However, PMA was clearly observed for the sample that underwent oxygen exposure. The PMA constant $K_u$ was roughly estimated to be $7.4 \times 10^6$ erg/cc. This value is larger than the $K_u$ for CoFeB/MgO systems. This result shows that oxygen exposure to CFS/MgO interfaces strongly enhances the PMA of the CFS layer beneath the MgO layer.

XPS analysis implied that MgO that had oxygen vacancies formed near the interface for the sample which did not undergo oxygen exposure, whereas MgO uniformly formed along the stacking direction by introducing the oxygen exposure process after deposition of the CFS layer. The oxygen terminated surface or slightly oxidized surface of the CFS layer could increase the Fe–O bonding, thus explaining the induced PMA.

The magnetic properties of the top CFS layer deposited at $T_S = 300°C$ were evaluated by subtracting the *M–H* curves of a sample with the Cr/Pd/CFS/MgO/Pd structure from those of a sample with the Cr/Pd/CFS/MgO/CFS/Pd structure (not shown here). The CFS layer deposited on the MgO barrier layer demonstrated isotropic magnetic anisotropy. Figure 2(a) shows AFM images of the CFS surface. The average surface roughness ($R_a$) of the top CFS layer that formed at $T_S$ = 300°C was 0.63 nm. This value is reasonable when the top CFS layer is grown with island-shaped morphologies resulting in isotropic magnetic properties. To obtain continuous layer growth with smooth surface morphologies, the fabrication conditions of the top CFS layer were refined as follows. $T_S$ was reduced to RT, and an additional annealing process at $T_A$ = 300°C was performed. Figure 2(b) shows an AFM image of the surface of the CFS layer formed via the refined process. $R_a$ reduced to 0.34 nm, indicating the formation of a continuous CFS film.

Figure 3 shows the magnetization characteristics of the Cr(40 nm)/Pd(50 nm)/CFS(0.6 nm)/MgO(2.0 nm)/CFS(0.6 nm)/Pd(10 nm) film when magnetic fields were applied along the out-of-plane direction. The steps for separated magnetization switching of the top and bottom CFS layers were clearly observed. The coercivity ($H_c$) of the top CFS layer was different from the bottom CFS layer, implying that the mechanism of PMA induced in the top CFS layer could be different from that in the bottom CFS layer.

Figure 4 shows the $\theta$–$2\theta$ scan XRD profiles for the Cr/Pd/CFS/MgO/CFS/Pd multilayers. The red pattern corresponds to the sample with the top CFS and Pd layers deposited at RT and subsequently annealed at 300°C whereas the blue pattern corresponds to the sample with the top CFS and Pd layers deposited at 300°C. The Pd capping layer was (111)-oriented for the sample fabricated with the refined fabrication conditions, but was not for the other sample. This (111) orientation of the Pd layer could explain the induced PMA in the top CFS because Co/Pd[23] and CMS/Pd interfaces[24] were previously reported to exhibit PMA when the Pd layers had an (111) orientation.

In summary, we demonstrated the addition of PMA to full-Heusler CFS alloy thin films using FTS, which is applicable to both the top and bottom magnetic electrodes of p-MTJs. A CFS layer with an MgO capping layer exhibiting PMA was successfully fabricated by forming a chemically uniform MgO layer along the growth direction. Additionally, PMA was clearly observed in the top CFS layer in the Cr/Pd/bottom-CFS/MgO(2.0 nm)/top-CFS/Pd multilayer, wherein the top CFS thin film was annealed at 300°C after deposition at RT to form a continuous CFS layer. In addition to the continuous layer growth of the films, the crystalline orientation alignment at the top CFS/Pd interface due to the appearance of the Pd (111) texture probably attributes to the origin of PMA at the top CFS layer. The process we introduced in this study, including oxygen exposure of the CFS surface or thinning of the CFS to 0.6 nm, may decrease TMR signals. Thus, further studies including magnet transport measurements are needed.

**ACKNOWLEDGMENTS**

A part of this study was supported by semiconductor technology academic research center. XPS measurements were performed in Yano group, Tokyo Institute of Technology, Japan.

Figure Captions

Figure 1. (a) Stacking structure. In-plane (blue) and out-of-plane (red) $M$–$H$ loops of CFS(0.6 nm)/MgO(2.0 nm) bilayer samples prepared (b) without and (c) with oxygen exposure after deposition of CFS layers.

Figure 2. AFM images of the surface of (a) the top CFS layer deposited at $T_S = 300°C$ and (b) at $T_S = $ RT and subsequently annealed at $T_A = 300°C$.

Figure 3. Out-of-plane $M$–$H$ loop of Cr(40 nm)/Pd(50 nm)/CFS(0.6 nm)/MgO(2.0 nm)/CFS(0.6 nm)/Pd(10 nm). The inset shows the sample structure.

Figure 4. $\theta$–$2\theta$ scan XRD profiles of the films with Cr/Pd/CFS/MgO/CFS/Pd structures. Red and blue patterns correspond to the sample in which the top CFS and Pd layers were deposited at $T_S = $ RT and then subsequently annealed at $T_A = 300°C$ and the sample in which the top CFS and Pd layers were deposited at $T_S = 300°C$, respectively.

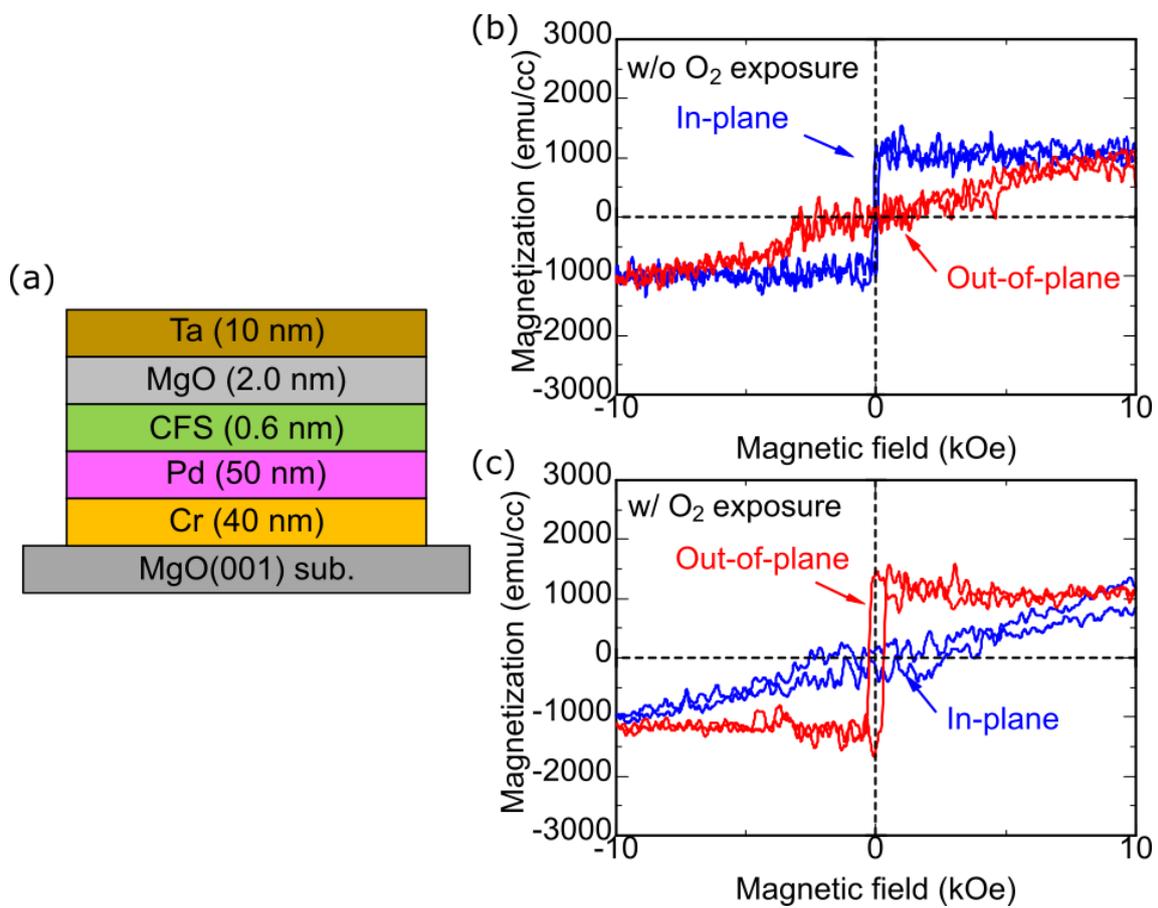

**Figure 1**

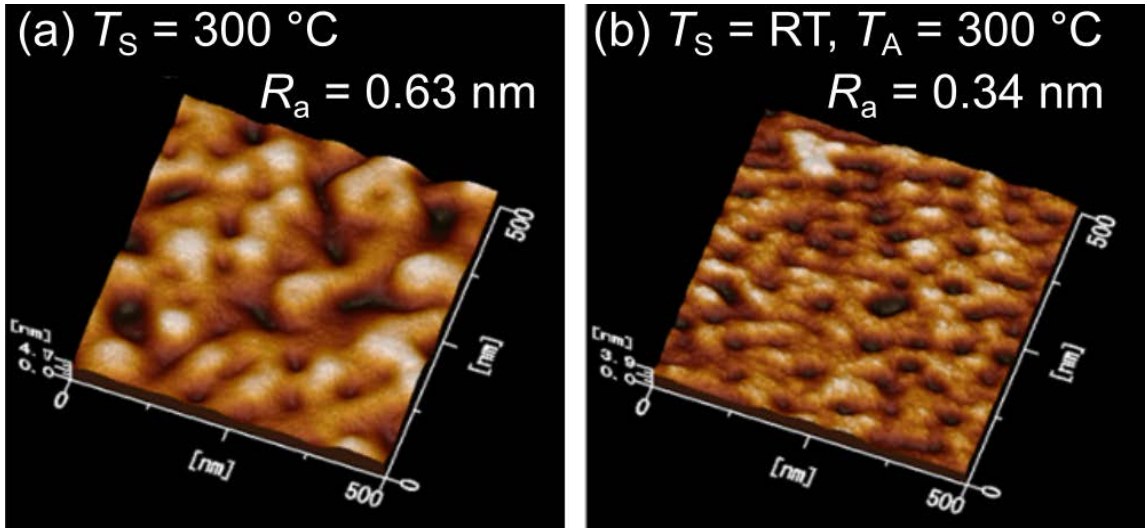

**Figure 2**

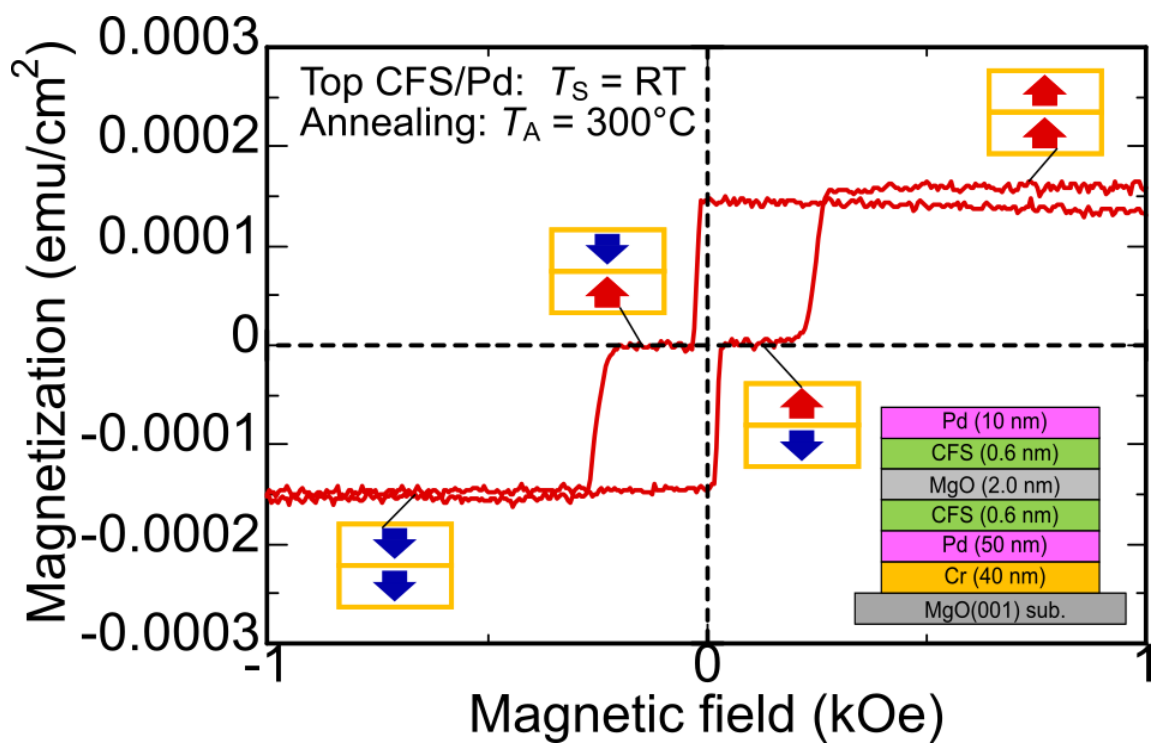

**Figure 3**

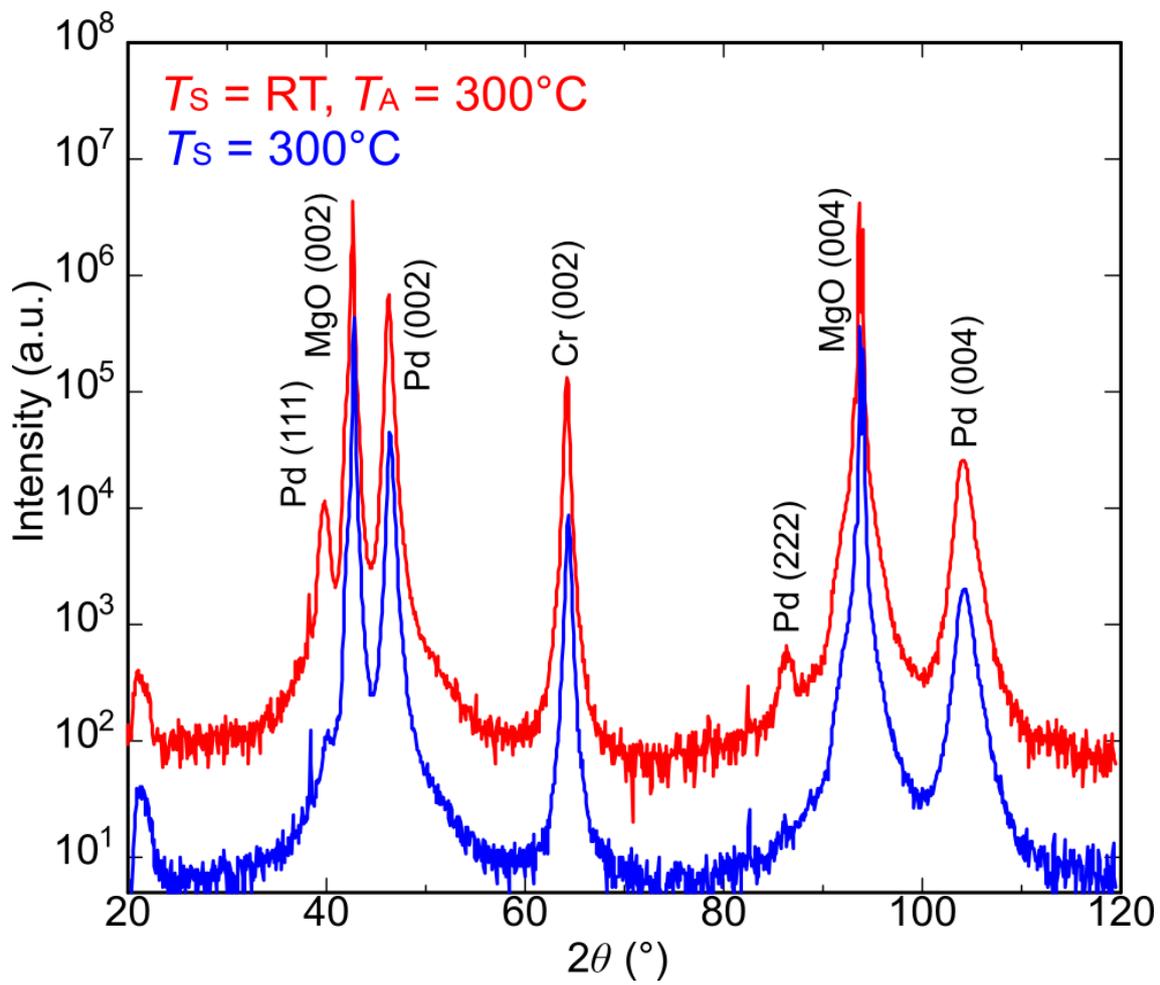

**Figure 4**